\documentclass[article,12pt,onecolumn,groupedaddress]{revtex4}

\usepackage{latexsym}
\usepackage{times}
\usepackage{graphicx}
\usepackage{hyperref}
\usepackage{epsfig}

\begin{document}
\thispagestyle{empty}
\pagestyle{empty}

\bibliographystyle{apsrev}

\title{Ground-Plane Quasi-Cloaking for Free Space}

\author{Efthymios Kallos}
\email[]{themos.kallos@elec.qmul.ac.uk}
\affiliation{School of Electronic Engineering and Computer Science, Queen Mary University of London, Mile End Road, London, E1 4NS, United Kingdom}
\author{Christos Argyropoulos}
\affiliation{School of Electronic Engineering and Computer Science, Queen Mary University of London, Mile End Road, London, E1 4NS, United Kingdom}
\author{Yang Hao}
\email[]{yang.hao@elec.qmul.ac.uk}
\affiliation{School of Electronic Engineering and Computer Science, Queen Mary University of London, Mile End Road, London, E1 4NS, United Kingdom}

\date{\today}

\begin{abstract}
Ground-plane cloak designs are presented, which minimize scattering of electromagnetic radiation from metallic objects in the visible spectrum. It is showed that simplified ground-plane cloaks made from only a few blocks of all-dielectric isotropic materials, either embedded in a background medium or in free space, can provide considerable cloaking performance while maintaining their broadband nature. A design which operates isolated in free space that cloaks radiation originating from a specified direction is also analyzed. These schemes should be much easier to be demonstrated experimentally compared to full designs.
\end{abstract}


\maketitle

Transformation electromagnetics allows the manipulation of light in unprecedented ways, such as the design of invisibility cloaks \cite{Pendry06-trans}. Most notably, a dispersive approximate cylindrical cloak that can be placed in free space was constructed using metamaterial-based resonating structures, and was demonstrated to work in the microwave regime \cite{Schurig05-cloaking}. However, due to causality constraints \cite{Tretyakov07,Yao08-limitations}, the required material anisotropy of such designs can only be achieved through increased absorption \cite{Linden04-loss} and through sacrificing the bandwidth of operation \cite{Ruan07-sensitive}. In order to find a way around these issues, a different transformation that produces a design which transforms an object to a metal sheet was proposed \cite{li08}. In this scenario, the object is only cloaked when placed above a ground plane and the whole device needs to be embedded in a background material. However, it has the advantages that the material anisotropy can be so minimal that the cloak works well if the anisotropy is simply ignored. This approach was verified experimentally by constructing a broadband microwave cloak using metamaterial elements operating non-resonantly \cite{Liu09-carpet}. Still, the cloak demonstrated in \cite{Liu09-carpet} was built using more than 6,000 unique metamaterial cells, which might not be straightforwardly extended to an optical ground-plane cloak due to the required sub-nanometer tolerances \cite{Cai07-optical-metamaterials}.

In this paper finite-difference time-domain (FDTD) simulations \cite{Taflove05} are performed to test simplified broadband ground-plane cloak designs that operate in the optical frequencies, embedded either in a background material or in free space. These designs are considered as quasi-cloaks, in the sense that they are not aimed at providing perfect cloaking, but rather use the main physical principles of transformation electromagnetics in order to minimize the scattering signatures of metallic objects. First, it is showed that a simplified quasi-cloak consisting of only six unique blocks of conventional materials (with relative permittivities $\varepsilon>1$) can provide very good ground-plane cloaking, after ignoring the anisotropy introduced from the initial transformation. Second, by further ignoring dispersive permittivity values ($\varepsilon<1$), it is showed that a similar simplified ground-plane quasi-cloak designed to be embedded in free space minimizes scattering, without requiring a surrounding impedance-matched layer. The performance of the quasi-cloaks is confirmed by evaluating the spatial and spectral distributions of the scattered field energy. Finally, a simplified optical quasi-cloak is analyzed, that operates directionally in air but without requiring a ground surface, a design that works similar to the optical mirage effect. These cloaks are isotropic, broadband, in principle lossless,  and can be constructed with only a few all-dielectric blocks of conventional materials, which should pave the way to demonstrate them experimentally at optical frequencies.

\begin{figure}[t]
\centering
\includegraphics[width=6.0cm]{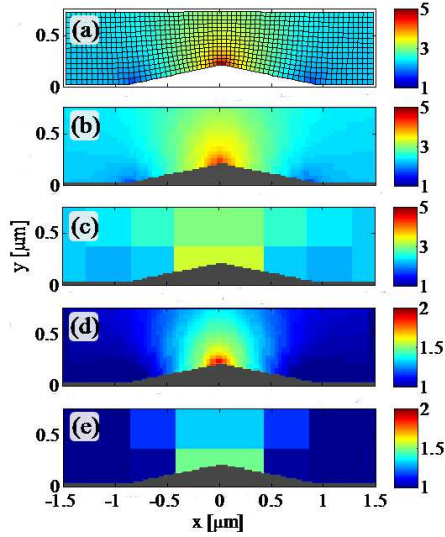}
\caption{Relative 2D permittivity maps for cloaking a triangular mettalic object placed over a ground plane. The cloaks in (a) - (c) are embedded in glass ($\varepsilon_{ref}=2.25$), while the cloaks in (d) - (e) are embedded in free space ($\varepsilon_{ref}=1$). The colored bars indicate the relative permittivity values for each map. (a) Full non-orthogonal map consisting of $64\times15$ cells. (b) High-resolution sampled map consisting of $80\times20$ blocks. (c) Low resolution sampled map consisting of $6\times2$ blocks. (d) High resolution sampled map consisting of $80\times20$ blocks. (e) Low resolution sampled map consisting of $4\times2$ blocks.} \label{f1}
\end{figure}

A two-dimensional (2D) geometry is considered, where a metallic object is placed on a flat ground plane. Without loss of generality, the object to be cloaked has a triangular shape with a height of $0.25 \mu$m and base of $1.8 \mu$m, and it is initially embedded in a glass material with $\varepsilon_{ref}=2.25$. The cloak is placed around the object, covering a region $3 \mu$m wide and $0.75 \mu$m tall. The permittivity distribution inside the cloak is found by generating a map that consists of $64\times15$ non-orthogonal cells, which compresses the space occupied by the object inside the cloaked region \cite{li08}, as shown in Fig. \ref{f1}(a). Given the $2\times2$ covariant metric $g$ for each cell \cite{Thompson}, the relative permittivity of each block is found as $\varepsilon=\frac{\varepsilon_{ref}}{\sqrt{\det{g}}}$ . A suitable map that reduces the anisotropy of the cloak is found by minimizing the width of the distribution of the parameter $\sqrt{\frac{g_{xy}g_{yx}}{g_{xx}g_{yy}}}$. For the map shown in Fig. \ref{f1}(a), $1.79\leq\varepsilon\leq4.70$ and the maximum anisotropy is $1.20$. The latter value can be reduced more by choosing to cloak an object with less sharp corners, however it is preferred here because the cloak can be constructed in a simple manner using straight cuts.

\begin{figure}[t]
\centering
\includegraphics[width=9.0cm]{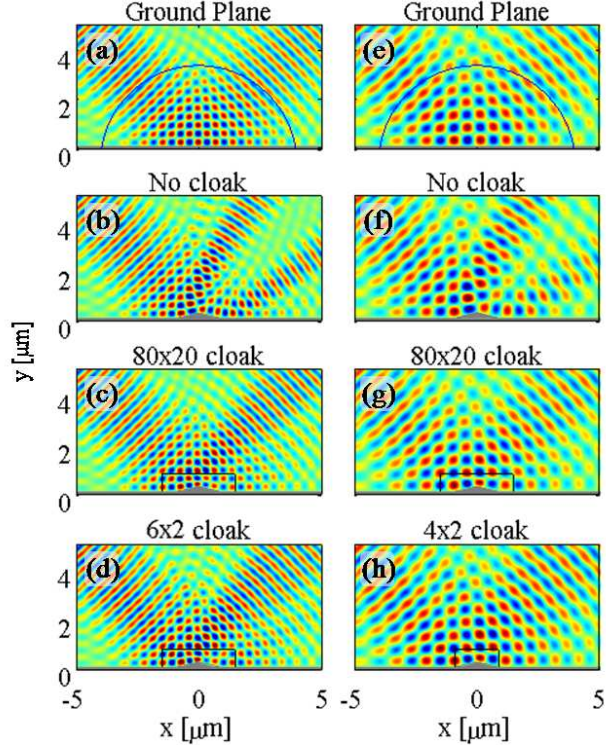}
\caption{Electric field amplitude distribution for a $2.4 \mu$m wide, $400$ THz Gaussian pulse impinging at $45^{\circ}$ angle with respect to the normal to a conductive plane. The location of the cloak and the object are outlined. A semi-circular curve with $4 \mu$m radius is also drawn for reference. (a)-(d): The background material is glass. (e)-(h): The background material is free space. (a),(e): Reflection off a flat plane. (b),(f): Scattering from a triangular metallic object. (c),(g): Cloaks comprising of $80\times20$ blocks cover the object. (d),(h): Simplified cloaks comprising of $6\times2$ (d) and $4\times2$ (h) blocks, respectively, cover the object.} \label{f2}
\end{figure}

Next, an orthogonal grid generated by recursive division of cartesian cells is used to sample the original permittivity distribution of Fig. \ref{f1}(a), as shown in Fig. \ref{f1}(b). The sampled map consists of $80\times20$ square blocks that have dimension equal to $0.0375 \mu$m. In addition, a low-resolution sampled map (quasi-cloak) is generated, consisting of $6\times2$ blocks that have dimensions $0.4285 \mu$m by $0.3750 \mu$m, as shown in Fig. \ref{f1}(c). Some of the blocks are truncated to fit around the object. The latter quasi-cloak has $2.18\leq\varepsilon\leq3.30$. From the point of view of an impinging electromagnetic wave, these two cloaks should behave similarly if its wavelength is not much smaller than the sizes of the blocks that consist the cloaks. In order to verify this claim, the performance of both cloaks is measured by using FDTD simulations with a resolution of $0.017 \mu$m, assuming a $2.4 \mu$m wide (full-width at half-maximum, FWHM), $400$ THz (in free space) TM Gaussian beam is incident at a $45^{\circ}$ angle onto the object (from the $x<0$ region). For this beam the free space wavelength is $\lambda_0=0.750 \mu$m. The simulation results are shown in Fig. \ref{f2}(a)-(d), when the anisotropy is ignored through setting the relative permeability of the device to $\mu=1$.

The reflection off a flat ground surface is shown first in Fig. \ref{f2}(a) for reference. The scattering off the aforementioned triangular metal object is shown in Fig. \ref{f2}(b), where two distinct energy side lobes are observed. When the high-resolution cloak of Fig. \ref{f1}(b) is placed around the object though, the reflection pattern of Fig. \ref{f1}(a) is mostly restored and a single beam is observed again near $45^{\circ}$, as shown in Fig. \ref{f2}(c). Any imperfections from the ideal reflected pattern are attributed to the ignored anisotropy, and can be mediated by using a larger cloaked region and/or a smoother object. Next, when the low-resolution quasi-cloak of Fig. \ref{f1}(c) is placed around the object, we observe in Fig. \ref{f2}(d) that the cloaking performance is almost identical to the performance of the high-resolution cloak (Fig. \ref{f2}(c)), thus verifying that ground-plane quasi-cloaking can be achieved with a very simple structure.

Subsequently, the performance of a ground-plane cloak embedded in free space is evaluated. A map similar to the one presented in Fig. \ref{f1}(a) is generated, with the difference that it is surrounded by free space with $\varepsilon_{ref}=1$. Unavoidably, the transformation generates cells near the base corners of the triangular object that correspond to permittivity values that are smaller than the background permittivity $\varepsilon_{ref}$, with a minimum value equal to $\varepsilon=0.8$. As it will be shown in detail later, since these regions are relatively small compared to the total size of the cloak and compared to the incident wavelength, they are not expected to impact the cloaking performance significantly. Thus, a sampled high-resolution cloak is generated for free space, consisting of $80\times20$ blocks, and any smaller than unity values of the permittivity are set to one, as shown in Fig. \ref{f1}(d). In addition, a low-resolution quasi-cloak is obtained by sampling the latter high-resolution permittivity map. This quasi-cloak is shown in Fig. \ref{f1}(e) and consists of $4\times2$ blocks with the following relative permittivity values in the $x<0$ domain (left to right, top to bottom): [1.17, 1.30, 1.02, 1.47]. The quasi-cloak is symmetric around $x=0$. The simulation results, when the previously described Gaussian beam impinges on this quasi-cloak, are shown in Fig. \ref{f2}(e)-(h), where it is observed that the field pattern of the cloaked object is very similar when either the low-resolution map (Fig. \ref{f2}(h)) or the high-resolution map (Fig. \ref{f2}(g)) are used. Again, the reflected beam pattern of the flat surface in free space (Fig. \ref{f2}(e)) is mostly restored from the two-lobe scattering pattern of the bare object (Fig. \ref{f2}(f)).

In order to quantify the performance and also verify the broadband cloaking capabilities of the ground-plane free space quasi-cloak of Fig. \ref{f1}(e), a $2.4 \mu$m wide, $4.7$ fs long, TM Gaussian pulse around $600$ THz is launched at $45^{\circ}$ (in the $x<0$ region) against the quasi-cloaked object. The total field energy crossing a semi-circular curve with $4 \mu$m radius centered at the object is recorded (in the $x>0$ region); this curve is drawn in Fig. \ref{f2}(a),(e). The pulse duration is chosen such that the frequency content of the pulse spreads over the whole visible spectrum: its FWHM is $\approx250$ THz.

\begin{figure}[t]
\centering
\includegraphics[width=8.0cm]{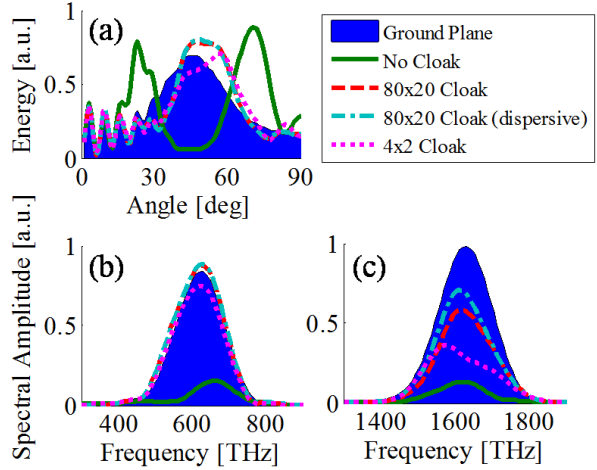}
\caption{(a) Angular distribution of the scattered field energy in free space when a $4.7$ fs long, $2.4 \mu$m wide, $600$ THz Gaussian pulse is incident. The patterns shown correspond to incidence on a flat plane, incidence on the metallic object placed on the plane, and incidence on the same object when covered with either the $80\times20$ (Fig. \ref{f1}(d)), the $80\times20$ dispersive, and the $4\times2$ (Fig. \ref{f1}(e)) quasi-cloaks. (b) The corresponding amplitudes of the frequency spectra of the scattered $600$ THz pulse as recorded at a $45^{\circ}$ angle. (c) The frequency spectra of a $1600$ THz scattered pulse in the same setup.} \label{f3}
\end{figure}

The angular distribution of the reflected energy in the $x>0$ region is shown in Fig. \ref{f3}(a); the angles are measured from the ground plane. When only the flat ground surface is present, the peak of the distribution is observed at a $45^{\circ}$ angle, as expected. When the metallic object is placed on top, however, two strong lobes are observed instead, at $23^{\circ}$ and $71^{\circ}$. When the quasi-cloak based on the $80\times20$ map (Fig. \ref{f2}(d)) is placed around the object, most of the scattered energy is now restored into a single lobe again around $49^{\circ}$. A similar single-lobe pattern is observed when the simplified sampled $4\times2$ quasi-cloak (Fig. \ref{f2}(e)) is utilized, as shown also in Fig. \ref{f3}(a), with only slight deterioration compared to the high-resolution cloak. The cloaking performance of simpler structures, consisting of fewer that eight blocks using this transformation map, is very limited. Note that the pattern observed for small angles (up to $\approx20^{\circ}$) in Fig. \ref{f3}(a) is a result of the interference between the incident and reflected parts of the pulse.

In addition, the effect of the dispersive values of the original map that were ignored is analyzed. A dispersive FDTD simulation \cite{Zhao08} with similar parameters is performed, with the difference that it includes the permittivity values $0.8\leq\varepsilon<1$ in the high-resolution $80\times20$ cloak, instead of setting that region to free space. The result of the angular distribution of the scattered energy for the dispersive cloak is also shown in Fig. \ref{f3}(a). Indeed, it is verified that the pattern is almost identical to the pattern of the non-dispersive $80\times20$ cloak at that frequency.

The frequency spectra of the $600$ THz scattered pulses are shown in Fig. \ref{f3}(b), by recording the electric field amplitude as a function of time on the semi-circular curve at a $45^{\circ}$ angle. The filled area indicates the spectrum of the reflected pulse when only the flat surface is present, which is used as a reference. We observe in Fig. \ref{f3}(b) that the frequency spectrum of the scattered pulse, when only the bare object is present on top of the ground plane, is severely distorted: both its amplitude is different compared to the reference spectrum, as well as its relative distribution within the spectrum. When the quasi-cloaks are covering the object, though, the spectrum of the original pulse at $45^{\circ}$ is almost fully recovered. Again, the dispersive sections of the cloak do not affect its performance when replaced by free space.

The strong broadband performance of the quasi-cloak is exhibited until the incident wavelength becomes much smaller than the dimensions of the cloak's block elements. This is illustrated by launching an identical $4.7$ fs long pulse at $1600$ THz frequency towards the cloak, and obtaining the reflected spectra as before, which are shown in Fig. \ref{f3}(c). It is observed that the simplified $4\times2$ quasi-cloak is not capable of restoring the spectrum as well as the $80\times20$ cloak, especially past $1550$ THz. In addition, the dispersive cloak (with its elements tuned to operate around the incident frequency) now exhibits improved performance compared to the all-dielectric one. The dispersive area is larger in terms of the incident wavelength for this higher frequency, thus introducing more error when replacing it with free space.

The results of Fig. \ref{f3} demonstrate that even though perfect optical cloaking is not achieved with a quasi-cloak, its cloaking performance is substantial, with the additional advantages that it is very simple to construct and it can be natively placed in free space. These two factors are obviously extremely favorable in a variety of cloaking applications, as experimental imperfections that are inevitably introduced when building more complicated structures, can be avoided. When designing a quasi-cloak, there is a tradeoff between the simplicity of the structure and the upper frequency of operation. While the broadband performance demonstrated here should be more than adequate for most applications, it can be improved when necessary by increasing the complexity of the structure.

The ground-plane quasi-cloaks discussed in this paper, so far, only work in the presence of a flat conducting surface. Next, we discuss a quasi-cloak design that can be placed isolated in free space, albeit working for a specified direction, as suggested in \cite{li08}. Starting with the quasi-cloak design of Fig. \ref{f1}(d), the ground plane is removed and the cloaked region along with the object is mirrored around $y=0$. As a result, a metallic diamond-shaped object is surrounded with a quasi-cloaking material consisting of $4\times4$ blocks. An incident electromagnetic wave would perceive a perfectly cloaked object as a collapsed strip of metal in the $y=0$ plane with - in principle - zero thickness. Suspended in free space, this device will scatter the impinging wave strongly for most incident angles, except when the wave is propagating parallel to the collapsed surface, i.e. along the $x$ direction in this setup, for which the object would be effectively cloaked.

\begin{figure}[t]
\centering
\includegraphics[width=9.0cm]{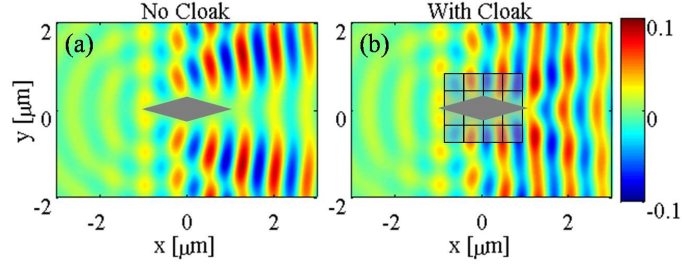}
\caption{Electric field amplitude distribution when a $12$ fs long, $400$ THz Gaussian pulse impinges on a diamond-shaped metallic object in free space, parallel to its long axis. (a) Bare object (b) Object covered with the proposed $4\times4$ simplified all-dielectric quasi-cloak.} \label{f4}
\end{figure}

In Fig. \ref{f4}(a) the electric field amplitude distribution when a $12$ fs long, $400$ THz TM Gaussian pulse is incident along the $+x$ direction on the bare object is shown, as calculated from FDTD simulations with the same discretization as before. A snapshot when the pulse has reached the $x>0$ region is depicted here. A very strong shadow is observed behind the object, along with a weak cylindrical scattering pattern in the $x<0$ region due to the sharp edge of the object at $x=-0.9 \mu$m. Subsequently, the aforementioned $4\times4$ quasi-cloak is placed around the object. We observe in Fig. \ref{f4}(b) that now there is no shadow left as the wavefronts are bending and recomposing on the back of the object, with only slight distortion. Along the line of propagation ($y=0$ axis), the total field energy restored behind the object is improved approximately 10 times compared to the non-cloaked object. Despite the weak reflections that remain in the $x<0$ region, this indicates strong cloaking performance which is on par with the performance of the approximate cylindrical dispersive cloak \cite{Schurig05-cloaking,Cummer06-cloak-simulations}. This scheme can be practical for a variety of cloaking applications, where the position of the observer is known in advance, e.g. for shielding objects placed in the line-of-sight of antennas or satellites. The scheme becomes increasingly favorable for quasi-cloaking objects in free space when considering its design simplicity, since only a few blocks of all-dielectric materials are required in order to achieve broadband lossless performance across the visible spectrum.

A physical explanation for the behavior of light waves in a ground-plane cloak, when the incidence occurs parallel to the ground plane (Fig. \ref{f4}(b)), is obtained when considering the well-known optical mirage effect \cite{Mirages}. In the latter case, the path of incoming light rays is bend towards increasing refractive index gradients, typically caused by air density differences in the atmosphere. In a similar manner, light traversing a ground-plane cloak is bend towards higher refractive indices, originating from the electromagnetic transformation. Three distinct stages are observed. First, as the wave reaches the first corner of the object (at $x=-0.9 \mu$m), the index gradient is increasing away from the surface (see also the map on Fig. \ref{f1}(a)), and thus the waves bend away from the object. Second, near the tip of the object (at the $x=0$ plane) the index gradient is reversed, causing the waves to bend back towards the object. Finally, due to the increasing index gradient away from the object near the second corner (at $x=+0.9 \mu$m), the waves are bend away from the object, and the original propagation parallel to the object's long axis is restored.

In conclusion, using FDTD simulations, it is demonstrated that ground-plane quasi-cloaks can be designed using relatively simple structures without significantly affecting their cloaking performance or bandwidth. Such a quasi-cloak designed to work in free space above a metallic surface, consisting of only eight all-dielectric blocks, can provide strong cloaking potential over the whole visible spectrum, as long as its features remain smaller than the wavelength of radiation. In addition, the quasi-cloak can operate without any flat surface present in free space for a specific angle of wave incidence, in analogy with the mirage effect. These designs should be straightforward to be practically implemented compared to previous ideas, e.g. by doping material blocks or by using non-resonant metamaterial cells.

The authors wish to thank Dr. Wei Song and Dr. Pavel Belov for creative feedback.

\bibliography{timaras2}
\bibliographystyle{apsrev}

\end{document}